\documentclass[aps,prb,twocolumn,groupedaddress]{revtex4}

\usepackage{graphicx}
\usepackage{dcolumn}
\usepackage{bm}

\begin{document}


\title{ Soft X-ray Magnetic Circular Dichroism of $c(2\times2)$ CuMn Ordered Surface Alloy }



\author{Akio Kimura}%
\email{akiok@hiroshima-u.ac.jp}
\author{Takafumi Kanbe}
\author{Tian Xie}
\author{Masaki Taniguchi}
\affiliation{%
Graduate School of Science, Hiroshima University, 1-3-1 Kagamiyama, Higashi-Hiroshima 739-8526, Japan\\
}%

\author{Shan Qiao}
\affiliation{
Hiroshima Synchrotron Radiation Center,Hiroshima University, 2-313 Kagamiyama, Higashi-Hiroshima 739-8526, Japan \\
}%

\author{Takayuki Muro}
\affiliation{
Japan Synchrotron Radiation Research Institute, Mikazuki, Hyogo 679-5143, Japan \\
}%

\author{Shin Imada, Shigemasa Suga}
\affiliation{
Graduate School of Engineering Science, Osaka University, 1-3 Machikaneyama, Toyonaka, Osaka 560-8531, Japan \\
}%



\begin{abstract}
Mn $2p$ soft X-ray absorption (XAS) spectroscopy excited with circularly polarized synchrotron radiation has been applied to a new class of material, $c(2\times2)CuMn$/Cu(001) two-dimensional ordered surface alloy.
A significant X-ray magnetic circular dichroism (XMCD) signal has been clearly observed at $T$=25K, indicating the existence of the ferromagnetic state under the external magnetic field of 1.4 Tesla.
The lineshape analyses of the XAS and XMCD spectra clearly show that the Mn $3d$ state is rather localized and has a high spin magnetic moment due to its half-filled character.
\end{abstract}

\pacs{73.20.-r, 73.22.-f}
\keywords{ two-dimensional surface alloy, ferromagnetism, soft X-ray absorption spectroscopy (XAS), soft X-ray magnetic circular dichroism (XMCD), multiplet structure }

\maketitle


\section{Introduction}
\begin{figure}
\includegraphics{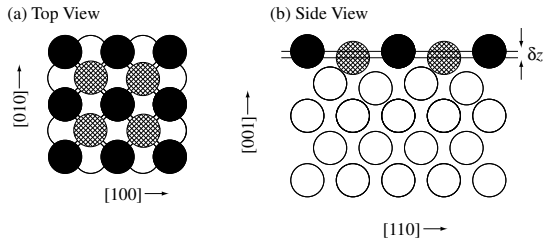}%
\caption{ Surface structure viewed from the top (a) and from the side (b) of $c(2\times2)CuMn$/Cu(001) 2D ordered surface alloy determined from the LEED $I-V$ analysis.
Here, the filled and hatched circles show the Mn and Cu atoms in the 1st layer and the open circle shows the Cu atoms located below 2nd layer.}
\end{figure}
It is known that Mn based two-dimensional (2D) ordered surface alloy can be formed on Cu(001) clean surface at a coverage of 0.5ML, where Mn and substrate atoms are alternatively placed forming a $c(2\times2)$ "checkerboard" structure as shown in Fig.1\cite{Wuttig93SS}.
A LEED $I-V$ measurement shows that $c(2\times2)CuMn$ surface alloy has a pronounced corrugation in which Mn atoms in the first layer are displaced outwards by $\delta z=0.30\pm0.02{\rm \AA}$, which is 17\% with respect to the atomic distance in the bulk\cite{Wuttig93SS}.
Such a remarkable atomic displacement is surprising because the other $c(2\times2)$ ordered surface alloy systems with non-magnetic elements, namely $c(2\times2)CuAu$/Cu(001) and $c(2\times2)CuPd$/Cu(001) show only small relaxations of 6\% and 1\%, respectively, instead of the larger atomic radius of Au(1.442\AA) and Pd(1.375\AA) compared to that of Mn(1.365\AA). 

A theoretical band structure calculation with FLAPW code predicts that the most stable magnetic state for $c(2\times2)CuMn$ orderd surface alloy is a ferromagnetic structure in the ground state\cite{Wuttig93PRL}.
The theory also explains that the observed large relaxation of the Mn atoms are derived from the magnetism\cite{Wuttig93PRL, Rader97PRB}.
The calculated magnetic moment of this surface alloy is 3.75$\mu_{\rm B}$, which is much larger than those of the bulk Mn crystal\cite{Rader97PRB}.

However, the experimental evidence of the ferromagnetic state of this surface alloy has not been obtained so far.
The reported result on a soft X-ray magnetic circular dichroism (XMCD) spectrum of the $c(2\times2)CuMn$/Cu(001) has shown no absorption intensity asymmetry at room temperature\cite{Brien93, Brien95}.
A spin resolved photoemission study at liquid nitrogen temperature has revealed almost negligible spin polarization of the photoelectron\cite{Rader97PRB}.
The lack of the experimental evidence possibly comes from the lower ferromagnetic transition temperature (Curie temperature) than liq.N$_{2}$ temperature as usually found in ultrathin films with a couple of $3d$ transiton metal monolayers\cite{Huang94}.
With this reason, we have tried to observe the soft X-ray magnetic circular dichroism (XMCD) spectra in the Mn $2p$ core absorption region at low temperature.

\section{Experimental}
Mn $2p$ core absorption spectroscopy (XAS) and X-ray magnetic circular dichroism (XMCD) spectra were measured at BL25SU of SPring-8.\cite{Suga01, Suga02, Saitoh01, Saitoh02}
Circularly polarized light was supplied from a twin-helical undulator, with which almost 100\% polarization was obtained at the peak of the first-harmonic radiation.
After having set the two undulators to opposite helicity, helicity reversal was realized by closing one undulator and fully opening the other.\cite{Saitoh01}
Mn $2p$ XAS spectra were measured by means of the total photoelectron yield method by directly detecting the sample current while changing the photon energy $h\nu$.
The measurement was performed in the Faraday geometry with both the incident light and the magnetization perpendicular to the sample surface.
We used two pairs of permanent dipole magnets with holes for passing the excitation light.
The external magnetic field of $\sim1.4T$ at the sample position was alternatingly applied by setting one of the two dipole magnets on the optical axis by means of a moter-driven linear feedthrough.
The XMCD spectra were taken for a fixed helicity of light by reversing the applied magnetic field at each $h\nu$.
In the present paper, the XMCD spectrum is defined as $\mu_{+}-\mu_{-}$, where $\mu_{+}$ and $\mu_{-}$ represent the absorption spectra for the direction of the majority spin parallel and antiparallel to the photon helicity, respectivery.
Clean surface was obtained by the repeated cycles of the Ar ion sputtering and annealing to 600$^{\circ}$C.
The cleanliness of the Cu(001) crystal surface was confirmed by the sharp $p(1\times1)$ LEED pattern and the absence of the C and O $LVV$ AES signals.
Manganese was evaporated from an electron beam evaporation source with a water cooling shroud at a rate of 0.2ML/min.
We finally observed very clear $c(2\times2)$ LEED pattern for 0.5ML Mn/Cu(001).
The temperature during the measurement was $\sim$25K.

\section{Results and Discussion}
The Mn $2p$ XAS and XMCD spectra of the $c(2\times2)CuMn$/Cu(001) are shown in Fig.2.
The XAS spectra are normalized by the incident photon flux.
We have observed several fine structures on each spin-orbit split components of the Mn $2p$ XAS spectrum.
We find two shoulder states at $\sim$1 and $\sim$2 eV higher energy of the $2p_{3/2}$ main peak.
Besides, the doublet peak structure has been found for the $2p_{1/2}$ component.
It is noted that the lineshape of the present Mn $2p$ XAS spectrum coincides well with the formerly reported result, where no XMCD has been observed at R.T.\cite{Brien93, Brien95}.
However, we have clearly observed the XMCD signal at $T$=25K as shown in the lower part of Fig.2, which indicates the existence of the long-range ferromagnetic order under the external magnetic field ($\sim$1.4T).
\begin{figure}
\includegraphics{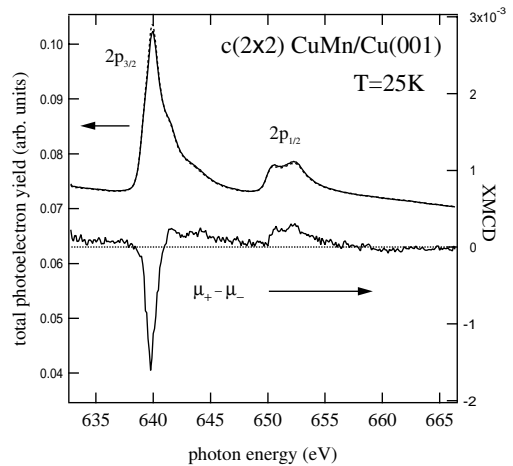}%
\caption{ Mn $2p$ XAS (upper) and XMCD (lower) spectra of $c(2\times2)CuMn$/Cu(001) 2D ordered surface alloy measured at $T$=25K.
The XAS spectra are normalized by the incident photon flux.
The XAS (XMCD) intensity scale is indicated at the left (right) axis.
}
\end{figure}
The observed asymmetry at the $2p_{\rm 3/2}$ XAS peak has been estimated to be only 2.4\%, which is quite small compared to the expected value for the fully magnetized sample($\sim57$\%).
The present XMCD spectrum shows the negative structure followed by the weaker positive structure with increasing $h\nu$ in the $2p_{3/2}$ core excitation region, whereas the double peak structure with the positive sign in the $2p_{1/2}$ region.
The overall lineshape of the XMCD spectrum is consistent with that of the other ordered surface alloy $c(2\times2)NiMn$/Ni(001)\cite{Brien95, Schmitz96}, in which the long-range ferromagnetic order in the $NiMn$ surface plane can be easily induced by the hybridization with the substrate atoms.
In the present data, the fine structures in the XMCD spectrum are better resolved compared to the case of $c(2\times2)NiMn$/Ni(001)\cite{Brien95, Schmitz96}, which may be due to the improved photon energy resolution and the localized nature of the Mn $3d$ electron in the $CuMn$ surface alloy.

Finally, we compare the experimental Mn $2p$ XAS and XMCD spectra with the calculated Mn $2p^{5}3d^{6}$ final state multiplets with assuming the $3d^{5}$ as the ground state configuration.
As shown in Fig.3, we find the excellent correspondence between the experimental XMCD spectrum as well as the XAS one, which shows that the observed several fine structures are derived from the multiplet effects.
This result clearly indicates that the Mn $3d$ electron has an almost half-filled electron nature leading to the high spin magnetic moment, which is consistent with the large exchange splitting of the Mn $3d$ states observed in the photoemission and inverse photoemission spectra\cite{Rader97PRB}.

The long-range ferromagnetic order under the external magnetic field has been proved from the present XMCD result.
However, it is still not clear if the ferromagnetic state is present even under zero magnetic field.
We can not exclude at present the possibility that the long-range ferromagnetic order is induced by a high external magnetic field for a (super-)paramagnetic ground state.
In order to clarify the existence of the ferromagnetic ground state, the extended measurement of the XMCD spectra as functions of temperature and applied magnetic field is highly desired.

\begin{figure}
\includegraphics{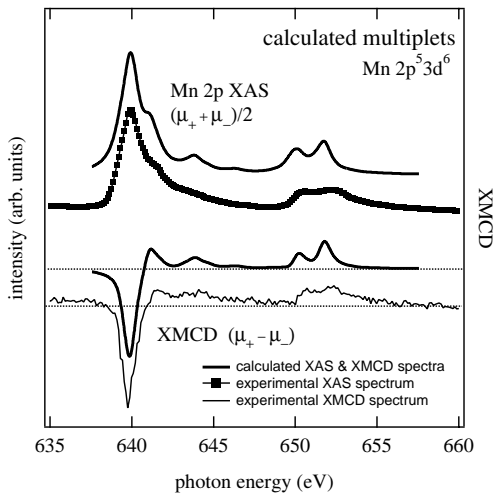}%
\caption{ Calculated Mn $2p$ XAS (upper) and XMCD (lower) spectra with Mn $2p^{5}3d^{6}$ final state multiplets (thick solid lines).
Experimental XAS (rectangle) and XMCD (thin solid line) spectra of $c(2\times2)CuMn$/Cu(001) are also shown for comparison.
The vertical sizes for the experimental spectra are scaled to the peak heights of the theoretical ones. 
}
\end{figure}

\section{Conclusion}
We have measured the Mn $2p$ XAS and XMCD spectra of $c(2\times2)CuMn$/Cu(001) 2D ordered surface alloy.
The observed XMCD clearly shows the existence of the long-range ferromagnetic order under the external magnetic field of 1.4T.
The lineshapes of the XAS and XMCD spectra are reproduced well by the calculated Mn $2p^{5}3d^{6}$ final state multiplets with $3d^{5}$ ground state, which suggests the half-filled nature of the Mn $3d$ electrons for this 2D ordered surface alloy.

\begin{acknowledgments}
The authors would like to thank Dr. Y. Saitoh of the Japan Atomic Energy Research Institute for a fine adjustment.
This work was done under the approval of the SPring-8 Advisory Committee (Proposal No. 2000A0492-NS1-np).
This work was supported by the Ministry of Education, Science, Sports and Culture.
\end{acknowledgments}

\end{document}